\documentclass[aps,preprint,showpacs,preprintnumbers,eqsecnum,amsmath,amssymb]{revtex4}

\usepackage{graphics,epsfig}
\usepackage{graphicx}
\usepackage{dcolumn}
\usepackage{bm}

\begin{document}

\title{First law of thermodynamics in IR Modified
H\v{o}rava-Lifshitz gravity}
\author{Mengjie Wang}
\author{ Jiliang {Jing}\footnote{Corresponding author,
Electronic address:
jljing@hunnu.edu.cn}}
\author{Chikun Ding}
\author{Songbai Chen}
 \affiliation{ Institute of Physics and
Department of Physics,
Hunan Normal University, Changsha, Hunan 410081, P. R. China \\
and
\\ Key Laboratory of Low Dimensional Quantum Structures and
Quantum Control of Ministry of Education, Hunan Normal University,
Changsha, Hunan 410081, P.R. China}

 \baselineskip=0.65 cm

\vspace*{0.2cm}
\begin{abstract}
\vspace*{0.2cm} We study the first law of thermodynamics in IR
modified H\v{o}rava-Lifshitz spacetime. Based on the
Bekenstein-Hawking entropy, we obtain the integral formula and the
differential formula of the first law of thermodynamics for the
Kehagias-Sfetsos black hole by treating $\omega$ as a new state
parameter and redefining a mass that is just equal to $M_{ADM}$
obtained by Myung\cite{YSM2} if we take $\alpha=3\pi/8$.

\end{abstract}

\vspace*{1.5cm}

\maketitle

\section{Introduction}

Many researchers are focused on black hole physics, and many
significant and interesting results have been achieved, including
Hawking radiation, black hole thermodynamics, and so on. In 1973,
Bardeen, Cater, and Hawking found that the integral formula for the
first law of black hole mechanics for a stationary axisymmetric
asymptotically flat black hole is given by \cite{Bardeen}
\begin{eqnarray}
M=\frac{\kappa}{4\pi}A+2\Omega_HJ_H+\frac{1}{4\pi}
\int_SR_a^b\xi_{(t)}^ad\Sigma_b, \label{101}
\end{eqnarray}
where $\kappa$, $A$, $\Omega_H$, $J_H$, $R_a^b$, $\xi_{(t)}^a$,
$d\Sigma_b$ are the surface gravity at the event horizon, the area
of the event horizon, angular velocity, angular momentum, Ricci
tensor, timelike Killing vector, and surface element, respectively.
Using Eq.(\ref{101}), they obtained the differential formula for the
first law of black hole mechanics \cite{Bardeen}
\begin{eqnarray}
\delta M=\frac{\kappa}{8\pi}\delta A+\Omega_H\delta
J_H+\int\Omega\delta d J+\int\overline{\mu}\delta d
N+\int{\overline{\theta}}\delta dS, \label{102}
\end{eqnarray}
where $\delta dN$ is the change in the number of particles crossing
$d\Sigma_b$, $\delta dS$ is the change in the entropy crossing
$d\Sigma_b$, $\overline{\mu}$ is the ``redshifted "chemical
potential and $\overline{\theta}$ is the red-shifted temperature.
Then Bekenstein \cite{Benkenstein} introduce the concept of
thermodynamics into black hole physics and Hawking \cite{Hawking}
proved that the black hole is indeed not entirely black and emits
radiation by using the quantum fields theory in curved spacetime.
Thus, the temperature and entropy of black holes are given by
\begin{eqnarray}
T=\frac{\kappa}{2\pi},\;\;\;\;\;\;\;\;\;\;\;\;
\;\;\;\;S=\frac{A}{4}.\label{103}
\end{eqnarray}

Recently, H\v{o}rava \cite{ho1,ho2,ho3} proposed a new class of
quantum gravity that is nonrelativistic and power-counting
renormalizable. It is a theory with higher spatial derivatives, and
the key property of this theory is the three-dimensional general
covariance and time reparameterization invariance. It is this
anisotropic rescaling that makes H\v{o}rava's theory power-counting
renormalizable. Therefore, a lot of attention has been focused on
this theory of gravity, and its cosmological applications have been
studied \cite{KS,cal,pia,gao,KK,RG,XYWR,DS}. Some static spherically
symmetric black hole solutions have been found in Ho\v{r}ava's
theory \cite{CY,LMP,CCO,JZT,EK,RM,CP}. The general IR vacuum has a
nonzero cosmological constant in Ho\v{r}ava's theory\cite{SMS}. In
order to get a Minkowsky vacuum in the IR region, one must add a new
term $\mu^4R^{(3)}$ in the action and take the
$\Lambda_{W}\rightarrow0$ limit for the cosmological constant
$\Lambda_{W}$. This does not change the UV properties of the theory,
but it alters the IR properties. Making use of such a modified
action, Kehagias and Sfetsos \cite{KS} obtained the asymptotic flat
spherically symmetric vacuum black hole solution(KS black hole).
This black hole behaves like the Reissner-Norstr\"{om} black hole
and has two event horizons. Moreover, the heat capacity is positive
for the small black hole, and it is negative for the large one. It
means that the small black hole is stable in the Ho\v{r}ava's
theory, which is quite different from that of the Schwarzschild
solution in Einstein's theory. The investigation of the quasinormal
modes of the massless scalar perturbations  shows that the
perturbations live longer in the IR modified H\v{o}rava-Lifshitz
spacetime \cite{CJ1,RAK1}. These results imply that distinct
differences exist between H\v{o}rava-Lifshitz's theory and
Einstein's gravity.

Because gravity theory, quantum theory, and statistical mechanics
are merged into black hole thermodynamics, it is believed that some
clues on quantum effects of gravity would be revealed in black hole
thermodynamics. Therefore, a lot of attention
\cite{MK,CCO1,CO,PW,WW} has been focused on black hole
thermodynamics for H\v{o}rava-Lifshitz gravity, and thermodynamics
for the KS black hole was also investigated and some peculiar
results were obtained \cite{ACAL,YSM}. The general procedure for
investigating KS black hole thermodynamics is to assume that the
first law of thermodynamics is  \cite{ACAL,YSM}
\begin{eqnarray}
dm=TdS, \label{104}
\end{eqnarray}
and the entropy is obtained as \cite{ACAL,YSM}
\begin{eqnarray}
S=\int\frac{d m}{T}+S_{0}=\pi
r_{+}^2+\frac{\pi}{\omega}\log(r_{+}^2)+S_{0}
=\frac{A}{4}+\frac{\pi}{\omega}\log(r_{+}^2)+S_{0}, \label{105}
\end{eqnarray}
with $S_0$ as an integration constant.

We have several comments regarding the above equations: (i) compared
with the Reissner-Nordstr\"om black hole, we know that
Eq.(\ref{104}) is flawed for the KS black hole, and it should be
modified with a work term. (ii) It is obvious that the integral
formula
\begin{eqnarray}
m= kTS, \label{106}
\end{eqnarray}
is not satisfied for the KS black hole, where $k$ is a proportional
constant (for example, $k=2$ for 4-dimensional Schwarzschild
spacetime and $k=(d-2)/(d-3)$ for the d-dimensional Schwarzschild
spacetime), and (iii) the expression for entropy of the KS black
hole is not consistent with Bekenstein-Hawking entropy, $S=A/4$.

In order to solve the above problems uniformly, motivated by
\cite{YS}, we find that $\omega$ in the KS black hole can be viewed
as a charge in some degree. By using the entropy of the KS black
hole, which is consistent with Bekenstein-Hawking entropy, and
redefining a new mass, the integral formula and differential formula
of the first law of thermodynamics for the KS black hole, which are
compatible with each other, can be obtained.

The remainder of this paper is organized as follows: In Sec. II, we
give a brief description of the solution in the IR modified
H\v{o}rava-Lifshitz black hole spacetime. In Sec. III, the integral
and differential form of the first law of thermodynamics for the KS
black hole are presented. In Sec. IV, the statistical entropy for
the KS black hole is studied. Finally, we summarize our conclusions
in the last section.

 \vspace*{0.4cm}
\section{Black hole in IR modified H\v{o}rava-Lifshitz gravity}

The general metric can be written in the following form in the
(3+1)-dimensional Arnowitt-Deser-Misner formalism:
\begin{eqnarray}
ds^2=-N^2dt^2+g_{ij}(dx^{i}+N^{j}dt)(dx^{j}+N^{j}dt), \label{201}
\end{eqnarray}
and its extrinsic curvature $K_{ij}$ is
\begin{eqnarray}
K_{ij}=\frac{1}{2N}\left(\dot{g}_{ij}-\nabla_i
N_j-\nabla_jN_i\right). \label{202}
\end{eqnarray}
In the H\v{o}rava theory, a modified action in the IR region is
given by \cite{KS}
\begin{eqnarray}
S_{HL}&=&\int dtd^3x \Big({\cal L}_0 + \tilde{{\cal L}}_1\Big),
\nonumber\\
{\cal L}_0 &=&
\sqrt{g}N\left\{\frac{2}{\kappa^2}(K_{ij}K^{ij}-\lambda
K^2)+\frac{\kappa^2\mu^2(\Lambda_W R^{(3)}
  -3\Lambda_W^2)}{8(1-3\lambda)}\right\}, \label{203}\\
  \tilde{{\cal L}}_1&=&
\sqrt{g}N\left\{\frac{\kappa^2\mu^2
(1-4\lambda)}{32(1-3\lambda)}(R^{(3)})^2 -\frac{\kappa^2}{2w^4}
\left(C_{ij} -\frac{\mu w^2}{2}R_{ij}^{(3)}\right) \left(C^{ij}
-\frac{\mu w^2}{2}R^{(3)ij}\right) +\mu^4R^{(3)} \right\},\nonumber
\\
\label{204}
\end{eqnarray}
where $\kappa^2$, $\lambda$, $\mu$, $w$, and $\Lambda _W$ are
constant parameters, $R^{(3)}$ and $R^{(3)}_{ij}$ are
three-dimensional spatial Ricci scalar and Ricci tensor. The Cotton
tensor $C_{ij}$ is
\begin{eqnarray}
 C^{ij}=\epsilon^{ik\ell}\nabla_k
\left(R^{(3)j}{}_\ell-\frac{1}{4}R^{(3)}
 \delta^j_\ell\right). \label{205}
\end{eqnarray}
Taking the $\Lambda_W\rightarrow0$ limit and letting $\lambda=1$, it
was found that the speed of light and the Newton constant are
described by the following relations \cite{KS}:
\begin{equation}
c^2= \frac{\kappa^2 \mu^4}{2}, \quad G = \frac{\kappa^2}{32 \pi c}.
\label{206}
\end{equation}

Considering a static and spherically symmetric background as
\begin{eqnarray}
ds^2=-N^2(r)dt^2+\frac{dr^2}{f(r)}+r^2 (d\theta^2+\sin^2\theta
d\phi^2 ), \label{207}
\end{eqnarray}
Eq.(\ref{204}) is changed into
\begin{eqnarray}
\tilde{{\cal
L}}_1=\sqrt{g}N\left\{\frac{3\kappa^2\mu^2}{64}(R^{(3)})^2
-\frac{\kappa^2\mu^2}{8}R^{(3)ij}R^{(3)}_{ij}+\mu^4R^{(3)}\right\},
\label{208}
\end{eqnarray}
where
\begin{eqnarray}
R^{(3)ij}R^{(3)}_{ij}=\frac{f'(r)^2}{r^2}+\frac{(-2+2f(r)+r
f'(r))^2}{2r^4},\;\;\;\;\;\;\;\;R^{(3)}=-\frac{2(-1+r
f(r)+f'(r))}{r^2}. \label{209}
\end{eqnarray}
Varying the action with N(r) and f(r) respectively, we then obtain
the KS black hole solution\cite{KS}
\begin{eqnarray}
N^2(r)=f(r)=1+\omega r^2-\sqrt{\omega^2 r^4+4\omega mr},\label{210}
\end{eqnarray}
where $m$ is an integration constant related to the black hole mass.

The parameter $m$ can be expressed by horizon radius $r_+$ as
\begin{eqnarray}
m=\frac{1+2\omega r_+^2}{4\omega r_+}. \label{211}
\end{eqnarray}
The outer and inner event horizon are given by
\begin{eqnarray}
r_{+}=m+\sqrt{m^2-\frac{1}{2\omega}},
\;\;\;\;\;\;r_{-}=m-\sqrt{m^2-\frac{1}{2\omega}}, \label{212}
\end{eqnarray}
with $m^2\geq\frac{1}{2\omega}$, and the extremal black hole should
be satisfied $m^2=\frac{1}{2\omega}$. The surface gravity is
\begin{eqnarray}
\kappa=\frac{f'(r_{+})}{2}=\frac{2\omega r_{+}^2-1}{4 r_{+}(1+\omega
r_{+}^2)}, \label{213}
\end{eqnarray}
and the corresponding temperature is given by
\begin{eqnarray}
T=\frac{\kappa}{2\pi}=\frac{2\omega r_{+}^2-1}{8\pi r_{+}(1+\omega
r_{+}^2)}. \label{214}
\end{eqnarray}

\section{first law of the KS black hole}
For an asymptotically flat black hole, Bardeen, Cater and Hawking
\cite{Bardeen} gave a differential geometric method to calculate the
integral formula and differential formula of black hole mechanics.
Because the KS black hole is an asymptotically flat black hole, we
will follow Bardeen, Cater and Hawking \cite{Bardeen} to obtain the
integral and differential formula for the KS black hole.

According to Eq.(\ref{210}), we know that there exist two Killing
vector fields, i.e., the timelike Killing vector field and the
spacelike Killing vector field, which are denoted by
 $\xi_{(t)}$ and $\xi_{(\varphi)}$, respectively.
 These Killing vector fields obey equations \cite{Bardeen}
\begin{eqnarray}
\nabla_b\xi_{(t)a}=\nabla_{[b}\xi_{(t)a]},\;\;\;\;\;
\nabla_b\xi_{(\varphi)a}=\nabla_{[b}\xi_{(\varphi)a]},\label{301}
\end{eqnarray}
\begin{eqnarray}
\xi_{(\varphi)}^{b}\nabla_b\xi_{(t)a}=\xi_{(t)}^{b}\nabla_b\xi_{
(\varphi)a},\label{302}
\end{eqnarray}
\begin{eqnarray}
\nabla^b\xi_{(t)b}^a=-R^{a}_{b}\xi_{(t)}^{b},\label{303}
\end{eqnarray}
\begin{eqnarray}
\nabla^b\xi_{(\varphi)b}^a=-R^{a}_{b}\xi_{(\varphi)}^{b},
\label{304}
\end{eqnarray}
where $\nabla_a$ represents a covariant derivative, and the square
brackets around indices imply antisymmetrization.

One can integrate Eq.(\ref{303}) over a hypersurface S and transfer
the volume on the left to an integral over a 2-surface $\partial S$
bounding $S$
\begin{eqnarray}
\int_{\partial S}\nabla^b\xi_{(t)}^ad\Sigma_{ab}=-\int_{S}R^a_b\xi_{
(t)}^bd\Sigma_a, \label{305}
\end{eqnarray}
where $d\Sigma_{ab}$ and $d\Sigma_a$ are the surface elements of
$\partial S$ and $S$, respectively. The boundary $\partial S$ of $S$
consists of $\partial S_{B}$ and a 2-surface $\partial S_{\infty}$
at infinity. Calculating the left expression in Eq.(\ref{305}) at
infinity, we have
\begin{eqnarray}
\int_{\partial S_{\infty}}\nabla^b\xi_{(t)}^ad\Sigma_{ab}=-4\pi m,
\label{306}
\end{eqnarray}
where m is the mass as measured from infinity.

Substituting Eq.(\ref{306}) into Eq.(\ref{305}), we have
\begin{eqnarray}
m=\frac{1}{4\pi}\int_{\partial
S_{B}}\nabla^b\xi_{(t)}^ad\Sigma_{ab}+\frac{1}{4\pi}
\int_{S}R^a_b\xi_{(t)}^bd\Sigma_a. \label{307}
\end{eqnarray}
Introducing the null vector $l^a$, which is equivalent to the
timelike Killing vector in our case
\begin{eqnarray}
l^a=\xi_{(t)}^a, \label{308}
\end{eqnarray}
using the surface gravity
\begin{eqnarray}
\kappa=l_{a;b}l^an^b, \label{309}
\end{eqnarray}
and the surface element of the event horizon $d\Sigma_{ab}$
\begin{eqnarray}
d\Sigma_{ab}=l_{[a} n_{b]}d A, \label{310}
\end{eqnarray}
Eq.(\ref{307}) can be rewritten as
\begin{eqnarray}
m=\frac{\kappa
A}{4\pi}+\frac{1}{4\pi}\int_{S}R^a_b\xi_{(t)}^bd\Sigma_a.
\label{311}
\end{eqnarray}
Because $\xi_{(t)}$ is a timelike Killing vector, we let
$\xi_{(t)}=(1,0,0,0)$ for the KS black hole. Therefore, the second
term in the right side of Eq.(\ref{311}) can be expressed as
\begin{eqnarray}
\frac{1}{4\pi}\int_{S}R^a_b\xi_{(t)}^bd\Sigma_a=
\frac{1}{4\pi}\int_{S}R^0_0\xi_{(t)}^0d\Sigma_0, \label{312}
\end{eqnarray}
with
\begin{eqnarray}
R^0_0=-\frac{3\omega^2\Big(2m^2+6\omega
mr^3+\omega^2r^6-4mr\sqrt{\omega^2r^4+4\omega mr}-\omega
r^4\sqrt{\omega^2r^4+4\omega mr}\Big)}{(\omega^2r^4+4\omega
mr)^\frac{3}{2}}. \label{313}
\end{eqnarray}
Substituting Eq.(\ref{313}) into Eq.(\ref{312}), we get
\begin{eqnarray}
\frac{1}{4\pi}\int_{S}R^a_b\xi_{(t)}^bd\Sigma_a=\frac{1+4\omega
r_{+}^2}{4\omega r_{+}(1+\omega r_{+}^2)}. \label{314}
\end{eqnarray}
Then substituting Eq.(\ref{314}) into Eq.({\ref{311}}), and
considering Eq.(\ref{103}), we have
\begin{eqnarray}
m=2TS+\frac{1+4\omega r_{+}^2}{4\omega r_{+}(1+\omega r_{+}^2)}.
\label{316}
\end{eqnarray}
Comparing Eq.(\ref{212}) with the outer and inner event horizon of
the Reissner-Nordstr\"om black hole
\begin{eqnarray}
r_{+}=m+\sqrt{m^2-Q^2},\;\;\;\;\;\;\;\;\;r_{-}=m-\sqrt{m^2-Q^2}\label{317}
\end{eqnarray}
and motivated by \cite{YS}, it is obvious that $\frac{1}{2\omega}$
is equivalent to $Q^2$, and this means that we could view
$\frac{1}{2\omega}$ as a charge in some degree. Therefore, we can
formally recast Eq.(\ref{316}) into
\begin{eqnarray}
M=2TS+V\frac{1}{\sqrt{2\omega}}, \label{318}
\end{eqnarray}
where $M$ is a new mass, and $V$ is the potential corresponding to
$\frac{1}{\sqrt{2\omega}}$. At the same time, the differential
formula of the first law should be taken in the following form:
\begin{eqnarray}
dM=TdS+Vd(\frac{1}{\sqrt{2\omega}}). \label{319}
\end{eqnarray}

According to the exact differential condition
\begin{eqnarray}
\frac{\partial}{\partial{\omega}}\Big(2\pi
Tr_+\Big)=\frac{\partial}{\partial{r_{+}}}\Big(-\frac{V}{2\sqrt{2}\omega^{\frac{3}{2}}}\Big),
\label{320}
\end{eqnarray}
and using Eqs.(\ref{319}) and (\ref{214}), we get the expressions
for $M$ and $V$
\begin{eqnarray}
M=\frac{r_{+}}{2}-\frac{3}{4\sqrt{\omega}}\arctan(r_{+}\sqrt{\omega})+h(\omega),\label{321}
\end{eqnarray}
\begin{eqnarray}
V=-\frac{3((1+\omega
r_{+}^2)\arctan(\sqrt{\omega}r_{+})-\sqrt{\omega}r_{+})}{2\sqrt{2}(1+\omega
r_{+}^2)}-g(\omega), \label{322}
\end{eqnarray}
where g($\omega$) and h($\omega$) are two integration parameters.
The relation between them is confined by Eq.(\ref{319}), i.e.,
\begin{eqnarray}
h'(\omega)=\frac{g(\omega)}{2\sqrt{2}\omega^{\frac{3}{2}}}.
\label{323}
\end{eqnarray}
Substituting Eqs. (\ref{321}) and (\ref{322}) into Eq. (\ref{318})
and together with Eq.(\ref{323}), we obtain
\begin{eqnarray}
h(\omega)=\frac{\alpha}{\sqrt{\omega}},\;\;\;\;\;\;g(\omega)=-\sqrt{2}\alpha,\label{324}
\end{eqnarray}
where $\alpha$ is an integration constant.

Then the expressions for $M$ and $V$ from Eqs. (\ref{321}),
(\ref{322}) and (\ref{324}) are
\begin{eqnarray}
M=\frac{r_{+}}{2}-\frac{3}{4\sqrt{\omega}}\arctan(r_{+}\sqrt{\omega})+\frac{\alpha}{\sqrt{\omega}},\label{325}
\end{eqnarray}
\begin{eqnarray}
V=\frac{3r_{+}\sqrt{\omega}+4\alpha(1+\omega r_{+}^2)-3(1+\omega
r_{+}^2)\arctan(\sqrt{\omega}r_{+})}{2\sqrt{2}(1+\omega
r_{+}^2)}.\label{326}
\end{eqnarray}
We should note that $\alpha>\frac{3}{4} \arctan{\frac{1}{\sqrt{2}}}
-\frac{1}{2\sqrt{2}}$  to keep $M$ is positive. We also note that
the expression for $M$ in Eq.(\ref{325}) is the same as the  mass
$M_{ADM}$ in Ref. \cite{YSM2} if we take $\alpha=\frac{3\pi}{8}$.

The above discussions show that the integral and differential
formula of the first law of thermodynamics for the KS black hole can
be expressed as Eqs.(\ref{318}) and (\ref{319}), and the expressions
for $M$ and $V$ are given by Eqs.(\ref{325}) and (\ref{326}).

\section{statistical entropy for the KS black hole}
The purpose of this section is to demonstrate that
Bekenstein-Hawking entropy also exists for the KS black hole. If we
recover $\hbar$, Eq.(\ref{105}) is changed into
\begin{eqnarray}
S=\int\frac{d m}{T}+S_{0}=\frac{\pi
r_{+}^2}{\hbar}+\frac{\pi}{\omega \hbar}\log(r_{+}^2)+S_{0}
=\frac{A}{4\hbar}+\frac{\pi}{\omega \hbar}\log(r_{+}^2)+S_{0}.
\label{ad1}
\end{eqnarray}
It is obvious that both the Bekenstein-Hawking term and the
logarithmic term have the same order of $\hbar$; therefore, it is
inappropriate for one to view the logarithmic term as a higher order
quantum correction.

In what follows, we strictly prove that the entropy obtained from
Eq.(\ref{ad1}) is incorrect and the Bekenstein-Hawking entropy,
$S=A/4$, also holds for the KS black hole by using the thin film
brick wall model, provided that only the order of $\hbar^{-1}$ is
considered. For simplicity, we only consider the massless case.

For massless particle, we have $P_\mu P^\mu=0$, i.e.,
\begin{eqnarray}
g^{tt}P_t^2+g^{rr}P_r^2+g^{\theta\theta}
P_\theta^2+g^{\varphi\varphi}P_\varphi^2=0. \label{401}
\end{eqnarray}
The module of the spatial component of the four-momentum is
\begin{eqnarray}
P^2\equiv P_j P^j=g^{rr}P_r^2+g^{\theta\theta}P_\theta^2+
g^{\varphi\varphi}P_\varphi^2=-g^{tt}P_t^2, \label{402}
\end{eqnarray}
and the number of the quantum state is
\begin{eqnarray}
\Gamma=\frac{1}{(2\pi\hbar)^3}\int{drd\theta d\varphi dP_r dP_\theta
dP_\varphi}. \label{403}
\end{eqnarray}
For convenience,  we set $ P_1^2=g^{rr}P_r^2, $
$P_2^2=g^{\theta\theta}P_\theta^2, $ $
P_3^2=g^{\varphi\varphi}P_\varphi^2,$ and take $
P^2=P_1^2+P_2^2+P_3^2. $ Then, Eq. (\ref{403}) becomes
\begin{eqnarray}
\Gamma &=&\frac{1}{(2\pi\hbar)^3}\int{drd\theta
d\varphi\frac{1}{\sqrt{g^{rr}g^{\theta\theta}
g^{\varphi\varphi}}}}\frac{4}{3}\pi
P^3\nonumber\\&=&\frac{1}{6\pi^2\hbar^3}
\int{\sqrt{g_{\theta\theta}g_{\varphi\varphi}g_{rr}}drd\theta
d\varphi(-g^{tt}P_t^2)^{\frac{3}{2}}}\nonumber
\\&=&\frac{(-P_t)^3}{6\pi^2\hbar^3}\int{\frac{\sqrt{-Det}}{g_{tt}^2}dr
d\theta d\varphi}, \label{404}
\end{eqnarray}
where $\sqrt{-Det}=\sqrt{-g_{tt}g_{rr}g_{\theta\theta}
g_{\varphi\varphi}}.$ The free energy is then given by
\begin{eqnarray}
F(\beta)&=&\frac{1}{\beta}\int{d\Gamma \ln(1-e^{-\beta
\omega})}\nonumber\\&=&-\int_{0}^{\infty}{\frac{\Gamma }
 {e^{\beta \omega}-1}d \omega}\nonumber\\&=&-\frac{1}{6\pi^2\hbar^3}
\int_{0}^{\infty}{\frac{(-P_t)^3}{e^{\beta \omega}-1}
d\omega}\int^{r_++\epsilon+\delta}_{r_++\epsilon}{\frac{\sqrt{-Det}}{g_{tt}^2}drd\theta
d\varphi}\nonumber\\&=&-\frac{2\pi^3}{45\beta^4}\int^{r_++\epsilon+\delta}_{r_{+}+\epsilon}\frac{r^2}{g_{tt}^2}dr.
\label{405}
\end{eqnarray}
From which we can get the entropy of the KS black hole
\begin{eqnarray}
S=\frac{8\pi^3}{45\beta(4\pi)^2}\frac{r_+^2\delta}{\epsilon(\epsilon+\delta)}=\frac{\pi
r_+^2}{90\beta}\frac{\delta}{\epsilon(\epsilon+\delta)}=\frac{A}{4},
\label{407}
\end{eqnarray}
where we let $\frac{\delta}{\epsilon(\epsilon+\delta)}=90\beta$.
Equation (\ref{407}) implies that the semiclassical entropy for the
KS black hole satisfies the Bekenstein-Hawking entropy.

\section{conclusion}
We have studied the first law of thermodynamics for the KS black
hole. If we assume that the differential formula of the first law of
thermodynamics is $dm=TdS$, some unsatisfactory results occur, i.e.,
the Bekenstein-Hawking entropy and the integral formula of the first
law of thermodynamics do not hold. By analogy with the
Reissner-Nordstr\"om black hole, we know that Eq.(\ref{104}) is
flawed for the KS black hole and it should be modified with a work
term. Based on Bekenstein-Hawking entropy, following the method
provided in Ref. \cite{Bardeen}, we obtain the integral formula
(\ref{318}) and the differential formula (\ref{319}) of the first
law of thermodynamics for the KS black hole by treating $\omega$ as
a new state parameter and redefining a new mass (\ref{325}).  The
new mass is just equal to $M_{ADM}$ in Ref. \cite{YSM2}, if we take
$\alpha=3\pi/8$.

\vspace*{0.2cm}
\begin{acknowledgments}
This work was supported by the National Natural Science Foundation
of China under Grant No. 10875040,  the key project of the National
Natural Science Foundation of China under Grant No 10935013, the
National Basic Research of China under Grant No. 2010CB833004,  the
Hunan Provincial Natural Science Foundation of China under Grant No.
08JJ3010, and the Construct Program of the National Key Discipline.
S. B. Chen's work was partially supported by the National Natural
Science Foundation of China under Grant No.10875041; the Scientific
Research Fund of Hunan Provincial Education Department Grant
No.07B043.
\end{acknowledgments}

\end{document}